\documentstyle[12pt,aaspp4]{article}	

\newcommand{\etal}{\mbox{\it et~al.\ }}

\newcommand{\msun}{\mbox{M$_{\odot}$~}}

\newcommand{\lsun}{\mbox{L$_{\odot}$}}
\newcommand{\kms}{\mbox{km s$^{-1}$}}

\def\deg      {{\ifmmode^\circ\else$^\circ$\fi} } 

\def\h2     {H$_2$}

\begin{document}

\title{The Host Galaxies of Optical Bright QSOs :
Molecular Gas in Z $\le$ 0.1 PG QSOs}

\author{N. Z. Scoville}
\affil{Astronomy Department, California Institute of Technology, Pasadena, CA 91125; nzs@astro.caltech.edu}

\author{D. T. Frayer }
\affil{SIRTF Science Center, California Institute of Technology, Pasadena, CA 91125; frayer@ipac.caltech.edu}

\author{E. Schinnerer}
\affil{Owens Valley Radio Observatory, California Institute of Technology, Pasadena, CA 91125; eschinne@zia.aoc.nrao.edu}

\author{M. Christopher}
\affil{Astronomy Department, California Institute of Technology, Pasadena, CA 91125; mc@astro.caltech.edu}

\begin{abstract}

We present results of a CO(1-0) line survey in a complete sample of 
12 low redshift (Z $\leq$ 0.1), optically bright QSOs (PG QSOs with M$_B \leq -23$ mag). Six new CO detections are reported here at levels 
exceeding I$_{CO}$ $\simeq$2 Jy \kms. Combined with three previously reported detections,
we find that 9 of the 12 QSOs have abundant, dense ISMs characteristic of 
late type galaxies. In all 9 of the detected QSOs, the derived molecular gas masses are
M$_{H2} \geq 1.0 \times 10^9$ \msun , with the most massive being 10$^{10}$ \msun ~(PG 0050+124 -- I Zw 1). In the three sources not yet detected in CO, the upper 
limits on the gas masses are $\sim 10^9$ \msun ~and thus we cannot rule out 
abundant ISMs even in these objects. Since our sample was chosen entirely 
on the basis of low redshift and optical luminosity (and  
not selected for strong infrared emission), we conclude that {\it the majority 
of luminous, low redshift QSOs have gas-rich host galaxies and  
therefore can not be normal elliptical galaxies}. 

\end{abstract}


\section{Introduction}

Despite numerous optical and near infrared 
studies, the nature of the host galaxies of luminous QSOs has remained a matter
of considerable debate since quasars were first discovered in 1964. The host galaxy type provides important
clues to the origin and fueling of the supermassive black holes 
, presumed to be the source of activity and luminosity in QSOs. 

Powerful radio galaxies and radio loud quasars appear
to be associated with massive elliptical galaxy hosts (L $>$ L$_*$; see \cite{dun01},
\cite{per01}). Both their light distributions and dominant old stellar 
populations are similar to massive E galaxies, consistent with the well-established empirical 
correlation that nuclear black holes have approximately 0.2 -- 0.5 \% of the bulge mass (\cite{mag98}, \cite{mer01}). Thus the black hole growth and feeding 
appears linked to the stellar bulge population which is dominant in elliptical galaxies.

On the other hand, evidence has also accumulated for an evolutionary 
link between merging ultra-luminous IR galaxies (ULIRGs) and UV/optical QSOs ( Sanders \etal 1988). The evidence includes : similar local space densities for ULIRGs and QSOs; FIR SEDs smoothly transitioning between the two classes (Sanders \etal 1989 \&
Neugebauer \etal 1986); AGN-like emission lines (\cite{vei99}) and significant point-like nuclei (less than 0.2 $ \arcsec $ -- \cite{sco00}) in 30-40\% of the ULIRGs ; and the association of both ULIRGs and some QSOs  with galactic interactions (\cite{mac84}, \cite{bah97}). In addition, at least
one radio-loud QSO, 3C48, shows strong Balmer absorption lines in its extended nebulosity indicating a luminous young stellar population (\cite{bor84}) and mm-CO emission (\cite{sco93}).

Perhaps the clearest discrimant between early and late type host galaxies is 
the host galaxy ISM, particularly the dense ISM probed by mm-CO emission. 
It is well established that normal E galaxies have molecular gas masses M$_{H2} \le 
10^8$ \msun ~while virtually all late type galaxies have M$_{H2} \ge 5\times 10^8$ and most have $\ge 10^9$ \msun ~(\cite{you91}; \cite{you89}; \cite{cas98}; \cite{geo01}). In contrast to optical/IR imaging,
the detection of CO from the host galaxy is not hampered by suppression of the 
bright quasar nucleus. We report here the results of an extensive survey 
to detect CO(1-0) emission associated with optically luminous QSOs in order to 
assess their host galaxy type.  

Our sample QSOs are from the Palomar-Green (PG) QSO survey including all
PG QSOs at z $\leq$ 0.1, M$_B$ $\leq -23$ mag and $\delta > 0$\deg (\cite{sch83}).  This redshift-limited QSO sample contains 12 objects. {\bf The sample was 
entirely selected based on optical brightness and is not biased 
toward IR-excess QSOs.} As IR-excess QSOs clearly have a large ISM dust mass to produce the IR excess (\cite{has00}; \cite{eva01}); choosing QSOs by 
IR selection obviously will not yield an unbiased sampling of luminous QSO 
host galaxies. Three sources have already been
observed and detected in CO (I~Zw~1 -- \cite{bar89}; \cite{sch98}, PG~1351+640 -- \cite{eva01}, PG~1440+356 -- \cite{eva01}). The previous 
detection experiments for these sources were initiated on the basis of their IR excesses and they are not reobserved here.  

\section{OVRO CO Observations and Calibration}

The PG QSOs listed in Table 1 were observed at $\lambda \sim$ 2.6 mm using the 
Owens Valley millimeter
array during the period 
of 1999 September -- 2001 May. The array consists of 6 10.4 m telescopes
and most of the data were obtained in the low resolution 
and equatorial configurations (as this was a detection experiment needing 
highest sensitivity). The typical SSB system temperatures were 250-400 K, corrected
for antenna and atmospheric losses.  Spectral resolution
was provided by a digital correlator configured with
four bands, each with 32 x 4 MHz channels, yielding a total velocity coverage $\Delta v \simeq 1300$ \kms ~at the redshifted CO(1-0) line.
The total integration time and spatial resolutions on the different sources varied depending on system scheduling and source declination (see Table 1).  
In addition to the spectral line correlator,
the continuum was measured in a 1 GHz bandwidth analog
correlator.  

Nearby radio loud quasars were used for gain and passband calibration. 
The data were calibrated using the standard Owens Valley array program 
(\cite{sco92}) and mapped using the NRAO AIPS package.

\section{CO Results}

\begin{figure}
\epsscale{0.9}
\plotone{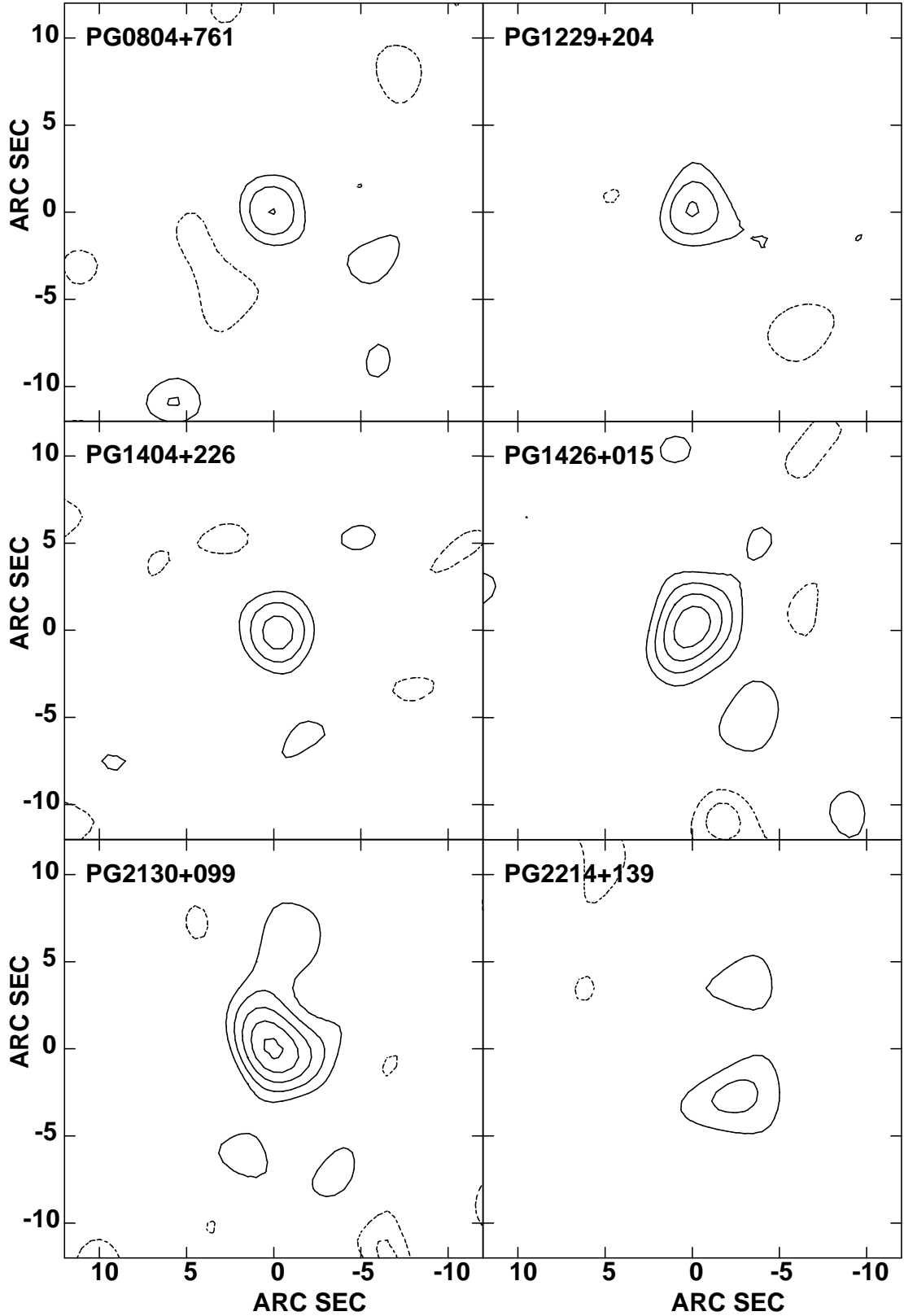}
\figcaption[Scoville_fig1.ps]{Maps of the integrated CO line flux are 
shown for the 6 newly detected objects. The origin of the 
coordinate offsets, given in Table 1 for each source, is the position of the 
optical QSO. The contours are -3, -2, 2, 3, 4, 5, 6 $\sigma$.
\label{co_maps}}
\end{figure}

Out of the nine new PG QSOs observed for CO emission, six were detected 
at $\geq$ 3 $\sigma$ levels. These detections along with upper limits for the remaining sources, are listed in Table 2. In all but one case, the CO(1-0) detections are at lower 
flux levels than any achieved previously (e.g. \cite{bar89}, \cite{eva01}).
Maps of the integrated CO flux are shown for the six detected sources in Fig.~\ref{co_maps}.
In no case was there detectable continuum emission, so the emission excesses are
clearly CO line emission. In all cases except PG 2214+139, the CO peak coincides 
within less than half the synthesized beamwidth of the optical QSO position;
in PG 2214+139, the displacement is 4\arcsec\ and thus the detection here needs further confirmation.  Lastly, we note that the non-detected sources 
have limits only a factor $\sim$ 2 below the detected sources; the non-detections 
should not therefore be interpreted as evidence that the molecular gas 
content of those sources is necessary very low. In PG 2130+099 the CO emission contours appear extended to the north of the nucleus at the 2 --3$\sigma$ level;
in all other sources, the CO is unresolved.  HST optical imaging (\cite{sur01}) for PG2130+099 shows the 
host galaxy to be an inclined spiral with major axis PA $\sim$ 45\deg. No other optical feature is seen in the north direction.

\subsection{H$_2$ Mass Estimates}

From the CO line flux integrated over the emission line (S$_{CO}\Delta v$),
we calculate the H$_2$ mass (M$_{H2}$) using 

$$ M_{H2}~~ =~~1.35\times 10^3~\alpha_{CO}~\left(S_{CO}\Delta v \over Jy~ \kms \right) \left(D^2_L \over Mpc^2 \right) \left(1 + z \right)^{-1} \msun  \eqno (1) $$

\noindent (\cite{sol92}). D$_L$ is the luminosity distance (with H$_0$ = 75 \kms ~Mpc$^{-1}$ and q$_0$ = 0.5). The CO-to-H$_2$ conversion factor is taken to be $\alpha_{CO} = 4$ \msun (K \kms ~pc$^2$)$^{-1}$ based on studies of molecular gas in the Galaxy (\cite{sco87}, \cite{str88}). In a few nearby ULIRGs, there is some evidence that the conversion factor could be
a factor of 2-3 lower (\cite{sco97}, \cite{dow98}) and if the lower 
values pertain to the objects observed here, the molecular gas masses 
would be lower by the same factor.

\subsection{Complete PG QSO Sample}

\begin{figure}
\epsscale{0.8}
\plotone{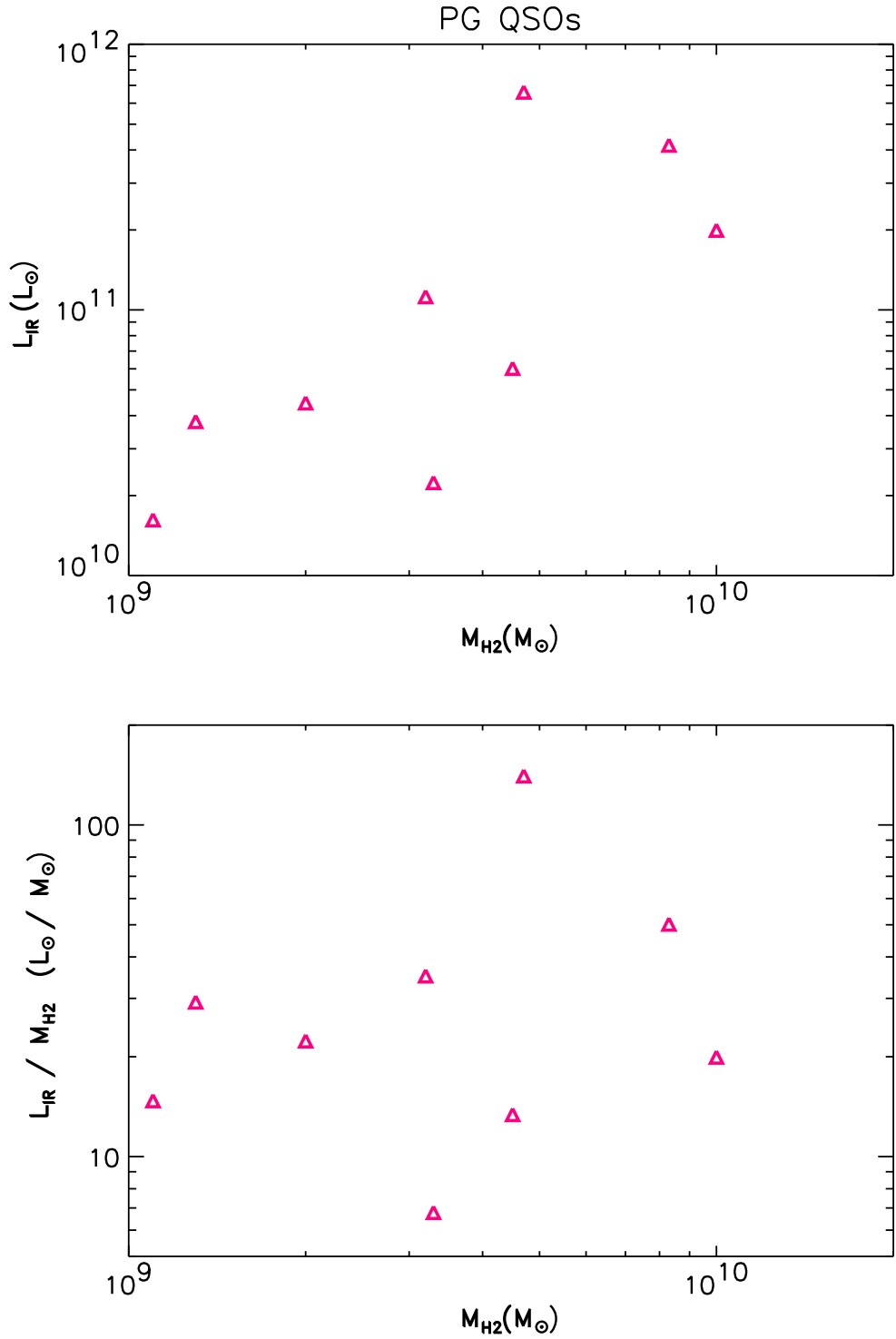}
\figcaption[Scoville_fig2.ps]{The far infrared luminosities (\cite{san89}) and 
luminosity-to-mass ratios are shown as a function of derived H$_2$ masses for the 9 Palomar-Green QSOs detected in CO. For a normal spiral galaxy like
the Milky Way, the luminosity-to-mass ratio is $\sim 4 $\lsun ~/ \msun 
and for the most active ULIRG galaxies the ratio is $\sim 200 $\lsun ~/ \msun.
\label{co_ir}}
\end{figure}

In Table 2, we list the H$_2$ mass estimates for all of the sources in our sample
including the three detected previously. The detected QSOs have masses in the 
range 1.6---10.0 $\times$ 10$^9$ \msun ~and the four non-detections have upper limits 
$\leq 1.5 \times 10^9$ \msun. Nine of the 12 sources have now been detected 
in CO and therefore have gas-rich ISMs, similar to those of late type 
galaxies or gas-rich merger remnants. In Fig.~\ref{co_ir}, the far infrared luminosities and 
luminosity-to-mass ratios are shown as a function of derived H$_2$ masses for the 9 Palomar-Green QSOs detected in CO. For a normal spiral galaxy like
the Milky Way, the luminosity-to-mass ratio is $\sim 4 $\lsun / \msun 
and for the most active ULIRG galaxies the ratio is $\sim 200 $\lsun / \msun;
the observed ratios for the PG QSOs are therefore entirely consistent
with those of gas-rich normal and starburst galaxies (10 -- 100 $\lsun / \msun$
, \cite{san91}). 

As a check on the mass estimates (and the CO-to-H$_2$ conversion factor), we
note that one of the sources in our sample, I Zw 1 (PG 0050+124),
was detected in the far infrared continuum with ISO (\cite{has00}). The derived 
dust mass from fitting the SED was 0.98 $\times 10^8$ \msun (\cite{has00}), compared with 
a H$_2$ mass of 10$^{10}$ \msun (Table 2). The implied gas-to-dust ratio of 
100 is entirely consistent with that obtained in the Galaxy and other nearby 
galaxies. None of the other sources in our sample were observed in the ISO study.

\section{Discussion}

It is important to stress
that these PG QSOs in our sample were {\bf not IR-selected and therefore, are truly
representative of normal optically luminous QSOs with M$_B \leq -23$ mag.} Two thirds
of the sample have molecular gas masses more than a factor 10 greater than 
any local {\bf normal} elliptical or S0 galaxy. The peculiar elliptical/radio galaxy Cen A does have a H$_2$ mass $\sim 10^9$ \msun~ but it is clear that this is the result of a recent merging with a gas-rich disk galaxy (\cite{mir99}). The clear implication is that 
these optically luminous QSOs are preferentially in gas-rich disk galaxies or 
gas-rich merger systems. 

Haas \etal (2000) have reached a similar conclusion based on ISO 
measurements of the far infrared continuum from a sample of 17 PG QSOs. 
Although similar in motivation, their results are not as easily interpreted
in terms of host galaxy type -- 10 of the 17 objects were detected with 
dust masses $\ge$ 10$^7$ \msun, corresponding to gas masses $\ge 10^9$ \msun
~but the sample was generated 'randomly' with a very large range of QSO 
properties and redshifts. None of their objects overlap with our sample.  Barvainis \& Ivison (2002) recently 
completed a submillimeter survey of 40 gravitationally lensed QSOs 
at Z = 1 -- 4.4 and detected 23 at 850 $\mu$m with M$_{dust} \sim 10^{7-8.4}$
\msun, implying gas masses of 10$^{9-10}$ \msun for standard gas-to-dust
ratios. Their important results show no differentiation between 
radio-loud and quiet AGN and clearly indicate that a significant fraction of 
high redshift AGNs have very dusty host galaxies. The results obtained
here demonstrate that these characteristics persist in the {\it majority} 
of optically luminous AGNs at the present epoch with perhaps a factor of 
2-3 lower ISM mass.   

How might one reconcile this conclusion with the studies using HST imaging which 
show light profiles in the hosts of the most luminous radio-loud and radio-quiet
quasars similar to those of giant elliptical galaxies (\cite{dun01})? The Dunlop \etal
sample had M$_B \leq -23.5$ mag; however, it would hardly seem reasonable that for a QSO sample with a lower cutoff 
only 0.5 mag more luminous, the host galaxy properties should change so fundamentally
from mostly disk/merger to old elliptical systems. We note that 3 of the galaxies
in our sample also have M$_B \leq -23.5$ mag. The Dunlop \etal result 
is also inconsistent with that of McLeod \& Rieke (1994) who obtained 
H-band light profiles for the host galaxies in 22 out of 24 local QSOs
with -24.1 $\leq$ M$_B \leq -23$. THey found that the majority were fit better by disk profiles 
than  r$^{1/4}$ laws. All of PG QSOs in our study were
observed by McLeod \& Rieke.

We also note that the r$^{1/4}$ law profile 
often used to identify elliptical galaxies is a very imprecise indicator
of galaxy type,
especially for merging systems which are certainly prevalent in the 
ULIRG samples (\cite{san96}). Scoville \etal (2000) found r$^{1/4}$ law profiles in 
over 40\% of the 24 luminous IR galaxies observed with NICMOS and a similar result was also obtained by Genzel \etal (2001). It is reasonable in 
the IR luminous galaxies to conclude that the bulge-like light profile was established by 
dynamical relaxation during the galaxy merging -- indeed this is also likely to 
be the cause of the same light profile in the massive ellipticals which have very 
likely undergone many merging events. The ULIRG light profiles have typical scale lengths a factor of 
a few smaller than the largest E galaxies -- very likely reflecting the larger mass
and larger number of merger events for the most massive E galaxies.   

Two possible scenarios linking the ULIRG and AGN phenomena
are : 1) that the abundant ISM which fuels the starburst
also feeds the central black hole accretion disk or 2) the post starburst stellar
population evolves rapidly with a high rate of mass-return to the ISM in the 
galactic nucleus, leading to sustained fueling of the black hole
(e.g. \cite{nor88}). Whether the entire QSO population had
precursor ULIRGs (implying that galactic merging is the
predominant formation mechanism for AGNs) or only a
small fraction of the QSOs are formed by {\it merging} of ISM-rich galaxies,
remains an open question. However, the molecular gas detections reported here
make it clear that the highest luminosity are
generally  not in 
{\it normal} (ie. gas-poor) ellipticals. Quite plausibly, the same process (galactic
merging) which can lead eventually to a relaxed, bulge stellar population also 
deposits large masses of ISM in the galactic nuclei to feed and 
buildup massive AGN. The connection between bulge light and central 
black hole mass (\cite{mag98}, \cite{mer01}) might then be indirect rather than a direct causal connection between the bulge stellar population 
and the buildup of the central massive black holes.

\vspace{5mm}

The Owens Valley millimeter array is supported by NSF grant AST 99-81546.

\clearpage

\begin{deluxetable}{l r r c r c c}
\tablenum{1}
\tablewidth{0pt}
\tablecaption{~~~~~~~~Summary of CO Observations}
\tablehead{ 
\colhead{~Source }&
\colhead{~RA (2000)\tablenotemark{a}}&
\colhead{~DEC (2000)\tablenotemark{a}}&
\colhead{~$cz$\tablenotemark{b}}&
\colhead{~Beam\tablenotemark{c}}& 
\colhead{~t$_{int}$\tablenotemark{d}}& 
\colhead{~rms\tablenotemark{e}~~~~~}\\
\colhead{}&
\colhead{}&
\colhead{}&
\colhead{\kms}& 
\colhead{\arcsec~$\times$~\arcsec}&
\colhead{hr}& 
\colhead{mJy beam$^{-1}$} 
}
\label{datalog}
\startdata
Detections : \\
PG\,0804+761 & 08:10:58.78 & 76:02:42.4 & 29918 & 4.3$\times$4.1 & 28 & 1.0 \\
PG\,1229+204 & 12:32:03.64 & 20:09:27.7 & 19207 & 4.7$\times$3.9 & 27 & 1.3 \\
PG\,1404+226 & 14:06:21.60 & 22:23:46.2 & 29337 & 5.1$\times$3.7 & 23 & 1.2 \\
PG\,1426+015 & 14:29:06.53 & 01:17:05.0 & 25990 & 4.5$\times$4.0 & 15 & 1.5 \\
PG\,2130+099 & 21:32:27.77 & 10:08:19.5 & 19010 & 4.9$\times$3.9 & 28 & 1.4 \\
PG\,2214+139 & 22:17:12.04 & 14:14:17.9 & 19517 & 6.1$\times$5.6 & 29 & 1.4 \\
\tableline
Non-detections : \\
PG\,0844+349 & 08:47:42.5 & 34:45:04.6 & 19097 & 4.5$\times$3.7 & 22 & 1.3 \\
PG\,1211+143 & 12:14:17.6 & 14:03:12.5 & 24253 & 3.6$\times$3.4 & 23 & 1.4 \\
PG\,1411+442 & 14:13:48.4 & 44:00:13.6 & 26861 & 4.9$\times$3.9 & 25 & 1.3 \\

\enddata
\tablenotetext{a}{The coordinates indicate the position obtained from fitting
a Gaussian to the detected emission. The offsets in Fig. 1 are relative to these positions. For the undetected QSOs, the optical QSO 
position is given.}
\tablenotetext{b}{The systemic velocity
is given as the heliocentric velocity using the optical convention
either denoting the line center for the detections or the middle of
the observed spectral band for the non-detections.\medskip}
\tablenotetext{c}{The fullwidth at half maximum of the
final synthesized beam.\smallskip}
\tablenotetext{d}{The total effective on-source integration time with 6 telescopes.}
\tablenotetext{e}{The rms derived from smoothed (330 \kms) maps for all
sources.\smallskip}
\end{deluxetable}

\begin{deluxetable}{lccccccc}
\tablenum{2}
\tablewidth{0pt}

\tablecaption{\large{\bf PG--QSOs (~z $< 0.1$ , M$_B < -23$~)}}
\tablehead{ 
\colhead{Name}
&\colhead{M$_B$\tablenotemark{a}}
&\colhead{log[L(IR)]\tablenotemark{a}}
&\colhead{z}
&\colhead{D$_L$~~\tablenotemark{b}}
&\colhead{S(CO)}
&\colhead{$\Delta$v$_{FWZI}$}
&\colhead{M$_{H2}$~~\tablenotemark{c}}\\
& (L$_{\odot}$) && (Mpc) & (Jy km/s) & km s$^{-1}$ & (10$^9$ M$_{\odot}$)
\\
}
\label{datalog}
\startdata
{\bf PG1351+640 } & -23.2 & 11.82 & 0.088 & 359 & 4.0$\pm$1.0 & 414 & 4.7~~\tablenotemark{d}  \\
{\bf PG1440+356 } & -23.4 & 11.62 & 0.078 & 318 & 9.0$\pm$1.5 & 600 & 8.3~~\tablenotemark{d}  \\
{\bf PG0050+124(I Zw 1) } & -23.5 & 11.30 & 0.061 & 248 & 18.0$\pm$2.0  & 650 & 10.2~~\tablenotemark{e}\\
{\bf PG1404+226 } & -23.1 & 11.05 & 0.098 & 401 &  2.0$\pm$0.5 & 468 & 3.2~~~\\
{\bf PG1426+015 } & -23.6 & 10.78 & 0.086 & 351 & 3.6$\pm$0.6 & 633 & 4.5~~~\\
PG1211+143 & -23.9 & 10.73 & 0.086 & 351 & $< 1.5$  & & $< 1.6$\\
{\bf PG2130+099 } & -23.2 & 10.65 & 0.061 & 248 &  4.3$\pm$0.6 & 487 & 2.0~~~\\
{\bf PG2214+139 } & -23.1 & 10.58 & 0.067 & 272 & $1.6$$\pm$0.5 & 488 & 1.3 \\
PG1411+442 & -23.7 & 10.52 & 0.089 & 363 &  $< 1.8$ & & $< 1.7$ \\
{\bf PG0804+761 } & -23.8 & 10.35 & 0.100 & 409 &  2.0$\pm$0.5 & 881 & 3.3~~~\\
PG0844+349 & -24.0 & 10.21 & 0.064 & 259 & $< 1.5$ & & $< 0.8$ \\
{\bf PG1229+204 }  & -23.3 & 10.21 & 0.064 & 259 & 2.4$\pm$0.6 & 266 & 1.1~~~\\
\enddata
\tablenotetext{a}{B magnitudes from Schmidt \& Green (1983); far infrared luminosities from Sanders et al. (1989).}
\tablenotetext{b}{\noindent Luminosity distance from $$  D_L = cH^{-1}_0q^{-2}_0 \left( z q_o +
\left(q_o -1\right)\left[\left(2 q_0 z +1\right)^{1/2} -1\right]\right) $$ Mpc assuming H$_0$ = 75 km s$^{-1}$ Mpc$^{-1}$ and q$_0$ = 0.5. }
\tablenotetext{c}{\noindent H$_2$ mass calculated using Eq. 1 with $\alpha$ = 4 M$_{\odot}$
[K km s$^{-1}$ pc$^2$]$^{-1}$. Upper limits to the line fluxes are 3$\sigma$.
PG2214+139 is detected only at the 3$\sigma$ level and therefore this 
detection should be treated somewhat tentatively. }
\tablenotetext{d}{Evans et al. 2000}
\tablenotetext{e}{Schinnerer et al. 1998,  Barvainis et al. 1989}
\end{deluxetable}

\clearpage

\newpage

\end{document}